\documentclass[prb,aps,twocolumn,showpacs,nofootinbib]{revtex4}
\usepackage{graphicx}
\usepackage{amsmath}
\usepackage{epstopdf}
\usepackage{amsbsy}
\newcommand{\bea}{\begin{eqnarray}}
\newcommand{\eea}{\end{eqnarray}}
\newcommand{\be}{\begin{equation}}
\newcommand{\ee}{\end{equation}}

\begin{document}


\title{Theory of Resistivity Upturns in Metallic Cuprates}
\author{W. Chen$^{1}$, Brian M. Andersen$^{2}$ and P. J. Hirschfeld$^{1}$}
\affiliation{$^{1}$Department of Physics, University of Florida, Gainesville,
FL 32611, USA \\
$^{2}$Nano-Science Center, Niels Bohr Institute, University of Copenhagen,\\
Universitetsparken 5, DK-2100 Copenhagen, Denmark
}
\begin{abstract}
We propose that the experimentally observed resistivity upturn of
cuprates at low temperatures may be explained by properly
accounting for the effects of disorder in a strongly correlated
metallic host. Within a calculation of the DC conductivity using
real-space diagonalization of a Hubbard model treated in an
inhomogeneous unrestricted Hartree-Fock approximation, we find
that correlations induce magnetic droplets around impurities, and
give rise to additional magnetic scattering which causes the
resistivity upturn.  A pseudogap in the density of states is shown
to enhance both the disorder-induced magnetic state and the
resistivity upturns. \vskip .4cm
\begin{center}(\today)\end{center}
\end{abstract}


\maketitle

\section{Introduction}

Exotic transport properties in metallic hole-doped cuprates
reflect their strongly correlated nature over a large part of
their phase diagram. In the vicinity of optimal doping, the
in-plane resistivity is found to be linear in temperature $T$,
with deviations from linear-$T$ power laws evolving on the
underdoped and overdoped sides.\cite{Chien91,Ito93}  As the
transition temperature $T_c$ is
 suppressed down to zero by a magnetic field, a resistivity that diverges
logarithmically at low temperatures (log-$T$) is  observed in
La$_{2-x}$Sr$_{x}$CuO$_{4}$ (LSCO)\cite{Ando95,Boebinger96} across
a wide range of doping. Resistivity ``upturns", increasing $\rho$
as temperature $T$ is decreased below a temperature $T_{min}$ have
been observed as well in
Ba$_{2}$Sr$_{2-x}$La$_{x}$CuO$_{6+\delta}$
(BSLCO)\cite{Hanaki01,Ono00} and sufficiently disordered and
underdoped  YBa$_{2}$Cu$_{3}$O$_{7-\delta}$ (YBCO) samples.
\cite{Fukuzumi96,Segawa99,Segawa01,Walker95,Rullier-Albenque00,Rullier-Albenque01,Rullier-Albenque03,Rullier-Albenque06,Rullier-Albenque07a,Rullier-Albenque07b}
Such upturns are frequently associated with a metal-insulator
transition as the system approaches its antiferromagnetic(AF)
parent compound (for a review, see Ref. \onlinecite{Alloul07}).
However, one should keep in mind that the explanation of these
resistivity upturns must include not only the intrinsic electronic
correlations present in the system, but also their interplay with
the external perturbations introduced to suppress
superconductivity.

Besides the suppression of $T_c$, it is known from inelastic
neutron scattering (INS), nuclear magnetic resonance (NMR), and
muon spin rotation ($\mu$SR) experiments that introducing disorder
and magnetic field can induce local magnetic order, reflecting the
coexistence of strong AF correlations  with superconductivity. For
instance, a strong signal centering at incommensurate positions
near $(\pi,\pi)$ has been observed in INS experiments in the
presence of a magnetic field\cite{Lake01,Lake02,Katano00,Khaykovich02}, indicating the
formation of AF order around vortices in LSCO; a smaller but
significant signal is also present in zero field.   Other neutron
scattering measurements have observed evidence of ordered static magnetism in intrinsically disordered cuprates, and shown that
systematic addition of disorder enhances this
effect.\cite{HKimura:2003,Suzuki98,Wakimoto01,Haug09}
NMR measurements have furthermore shown evidence that local
magnetic moments are induced around atomic scale defects such as
Zn substitutions of planar Cu, or defects produced by electron
irradiation.\cite{Alloul07,Ouazi04,Ouazi06,Bobroff99,Bobroff01}
The susceptibility of these induced moments shows a Curie-Weiss
behavior even though the impurity itself is nonmagnetic,
indicating their origin in the strong magnetic  correlations
present in the pure system. Finally,  $\mu$SR experiments have
shown that the Cu spins freeze in the underdoped superconducting
state, and eventually develop short range order at very low
temperatures in intrinsically disordered cuprates and even in the
much cleaner system YBCO if it is highly
underdoped.\cite{ChNiedermayer:1998,CPanagopoulos:2002,TAdachi:2004,SSanna:2004,CPanagopoulos:2005,RIMiller:2006}
  The relationship between ordinary
disorder and local magnetism in these and other experiments,  has
been reviewed in Ref. \onlinecite{Alloul07}, together with a
description of recent theoretical work.   Since these phenomena
are well established, a theory which seeks to account for the
transport anomalies should therefore also be capable of explaining
the formation of these local moments, as well as their ordering
behavior at different dopings and temperatures.

The logarithmic temperature dependence of the resistivity upturns
in a magnetic field has remained a mystery.     It is tempting to
associate these logs with the quantum corrections to the
conductivity found in weak localization
theory\cite{Altshuler79,Altshuler80,PALee85}. Indeed, in
electron-doped cuprates\cite{Fournier00}, where interaction
effects are thought to be weaker and disorder effects stronger, as
well as in overdoped cuprate samples\cite{Rullier-Albenque01},
good fits of the magnetoresistance data to weak localization
theory have been obtained. By contrast, elastic free paths in
hole-doped samples are much larger than the Fermi wavelength scale
required for weak localization effects; furthermore the
magnetoresistance has the wrong field dependence and typically
(but not always\cite{Hanaki01}) the wrong sign. A log-$T$ behavior
of the resistivity is also found in the theory of granular
systems,\cite{Beloborodov07} but evidence for granularity in the
conventional sense is weak or absent in the cuprate samples where
the upturns have been observed. Finally, it has been argued by
Alloul and others that the body of experimental results on
underdoped cuprates, specifically Zn-substituted and irradiation
damaged YBCO samples, is consistent with a one-impurity Kondo
picture, with conventional resistivity minimum. However there are
several inconsistencies associated with this approach, reviewed in
Ref. \onlinecite{Alloul07}.  We adopt here the alternate point of
view that the upturns observed in the underdoped, hole-doped
cuprates are manifestations of disorder in a Fermi liquid in the
presence of strong antiferromagnetic correlations.


A theory that can cover the anomalies of transport properties in
cuprates over a wide range of doping does not currently exist.
Recently, an attempt was made to treat disorder and interactions
in a model tailored to the cuprates by Kontani {\it et
al.}\cite{Kontani99,Kontani06,Kontani08} Within the
fluctuation-exchange (FLEX) approach and  certain approximations
regarding the impurity scattering processes and self-energy,
these authors had considerable success in reproducing resistivity
upturns observed in some cuprates in zero magnetic field. However, as a
 perturbative approximation it perforce
neglects certain self-energy and vertex correction diagrams;  in
addition, the physical content of the approximations made is not
always clear.

Here we focus on the optimally--  and slightly underdoped
cuprates, in the spirit of Kontani {\it et al.}\cite{Kontani08}, and assume the Fermi
liquid picture properly describes the electronic excitations in
the normal state.   We examine the following simple hypothesis
that connects the transport anomalies with the impurity induced
magnetization: the resistivity upturns are due to the extra
scattering associated with the correlation-induced magnetic
droplets which carry local moments.  Within a 2D single band
Hubbard model where interactions are treated in mean field but
disorder is treated exactly, we show that the resistivity
increases coincide with the conditions which enhance  impurity
induced magnetic moments.  
The present study focusses on the doping regime where static
moments, even in most strongly correlated LSCO, are paramagnetic
centers induced by the applied field.   Other recent studies
relevant to this phase have examined the more disordered, or more
correlated state where such magnetic droplets are spontaneously
formed around defects in zero field, and shown that they can
indeed affect macroscopic observables such as NMR, thermal
conductivity and superfluid density
$\kappa(T)$.\cite{Alvarez05,Andersen07,Atkinson07,Andersen06,Andersen08,Alvarez08}
The physical picture of the ground state, that of an inhomogeneous
mixture of AF droplets carrying net moments near the defect, is
quite similar in our case.  By working in the regime where moments
are smaller and the effect of the field is larger, however, we
hope to explain some of the observed puzzling aspects of the
magnetoresistance.  Since we consider relatively weak
correlations,  we explicitly confine ourselves to the doping
regions in each system under consideration where the resistivity
upturns first set in.  This means that we, within a RPA treatment
of the correlations, do not expect to be able to describe the true
MIT or log-$T$ behavior, but rather the leading perturbative
corrections to the high-$T$ behavior of the resistivity.   The
conditions in which positive correlations between impurity-induced
magnetization and transport anomalies can be found are examined, which confirm our hypothesis that the
enhancement of the scattering rate is due to an enlarged cross
sections associated with these induced moments. We  first examine
the case of optimal doping, and then  discuss the effect of
including a pseudogap in the density of states, which will allow
us to extend the model to lower dopings.

\section{Model Hamiltonian}

Since resistivity upturns are revealed after $T_{c}$ is suppressed
down to zero, the pairing correlation is ignored in describing the
normal state properties as a first approximation. We therefore
start with the two dimensional Hubbard model to describe the
CuO$_2$ plane
\begin{equation}
H=\sum_{ij\sigma}-t_{ij}c^{\dag}_{i\sigma}c_{j\sigma}
+\sum_{i\sigma}(\epsilon_{i\sigma}-\mu){\hat n}_{i\sigma}
+\sum_{i}U{\hat n}_{i\uparrow}{\hat n}_{i\downarrow}\;,
\label{Hubbard_bare_H}
\end{equation}
where $c_{i\sigma}$ is the electron operator at site $i$ with spin
$\sigma$, ${\hat n}_{i\sigma}=c^\dagger_{i\sigma}c_{i\sigma}$,
$t_{ij}=t,t'$ is the hopping amplitude between nearest-neighbor
($t$) and next-nearest-neighbor ($t'$) sites, and $U$ is the
onsite Coulomb repulsion. The external  perturbation due to
impurities and magnetic field is included in $\epsilon_{i\sigma}$
\begin{equation}
\epsilon_{i\sigma}=-\frac{1}{2}\sigma g\mu_{B}B
+\sum_{r}\delta_{ir}V_{imp}\;,
\end{equation}
where $V_{r}$ is the scattering potential produced by defects such as Zn substitution or electronic irradiation. The Zeeman term takes into account the spin-dependent energy shifts caused by the magnetic field with $\sigma=+/-$ for spin up/down respectively. We will denote $\frac{1}{2}g\mu_{B}B\equiv B$ in the figures presented below. A Hartree-Fock mean field decomposition is then adopted to the above Hamiltonian
\begin{eqnarray}
n_{i}&=&\langle {\hat n}_{i\uparrow}+{\hat n}_{i\downarrow} \rangle \nonumber \\
m_{i}&=&\langle {\hat n}_{i\uparrow}-{\hat n}_{i\downarrow} \rangle \;,
\label{Hubbard_op}
\end{eqnarray}
and gives rise to
\begin{eqnarray}
H&=&\sum_{ij\sigma}-t_{ij}c^{\dag}_{i\sigma}c_{j\sigma}
+\sum_{i\sigma}(\epsilon_{i\sigma}-\mu){\hat n}_{i\sigma}\nonumber \\
&+&\sum_{i\sigma}U\frac{n_{i}-\sigma m_{i}}{2}{\hat n}_{i\sigma}\;.
\label{Hubbard_MF_H}
\end{eqnarray}
By applying this mean-field Ansatz we certainly cannot study Mott
correlations in the regime $U \gg t$. We can, however, discuss
qualitatively the physics arising from the tendency to form
magnetic moments near impurities, as is appropriate near optimal
doping. To study the response of the system in a static electric
field, we calculate conductivity via linear response theory, where
we use the  current in the $x$-direction,
\begin{eqnarray}
J_{i}=J^{x}_{i}&=&it\sum_{\sigma} ({\hat c}^{\dag}_{i+x\sigma}{\hat c}_{i\sigma}-{\hat c}^{\dag}_{i\sigma}{\hat c}_{i+x\sigma})\nonumber\\
&+&it^{\prime}\sum_{\sigma} ({\hat c}^{\dag}_{i+x\pm y \sigma}{\hat c}_{i\sigma}-{\hat c}^{\dag}_{i\sigma}{\hat c}_{i+x\pm y\sigma})\;,
\label{tight_binding_current}
\end{eqnarray}
and express the site-dependent current-current correlation function in terms of eigenstates and eigenenergies
\begin{eqnarray}
\pi_{ij}(t)&=&-i\Theta(t)\left\langle \left[ J_i(t),J_j(0)\right] \right\rangle\;,\nonumber\\
\pi_{ij}(\omega)&=&\sum_{n,m}\left\langle n|J_i  |m\right\rangle \left\langle m|J_j |n\right\rangle \frac{f(E_n)-f(E_m)}{\omega+E_n -E_m +i\eta}\;,\nonumber\\
\sigma_{i}&=&\sum_{j}-\lim_{\omega\rightarrow 0}\left\lbrace\frac{Im\left(\pi_{ij}(\omega)\right)}{\omega} \right\rbrace \nonumber\\
&=&\pi\sum_{j}\sum_{n,m}\left\langle n|J_i |m\right\rangle \left\langle m|J_j |n\right\rangle F(E_n ,E_m)\;.
\label{jj_cor_fn}
\end{eqnarray}
The global conductivity $\sigma$ is realized by averaging
$\sigma_{i}$ over the whole sample with a proper normalization.
The function $F(E_n ,E_m)$ is symmetric under exchange of
$E_n\leftrightarrow E_m$.\cite{Takigawa02}
The resistivity $\rho$ is then given by the inverse of $\sigma$,
and is plotted in units of 2D resistivity $\hbar/e^{2}$. One can
also convert it into a 3D resistivity for materials such as YBCO,
in which one assumes two conducting planes per unit cell and gives
$\rho$ as 3D resistivity in units of $241\mu\Omega\cdot$cm.  Note
that this procedure gives us only the resistivity part due
to impurity scattering; since the Hamiltonian (\ref{Hubbard_MF_H})
is decoupled at the Hartree level, the inelastic processes which
lead to, e.g. the linear resistivity at optimal doping are not
treated.  We calculate therefore only the low-$T$ part due to
elastic scattering.

The proper choice of system size in simulating Eq.
(\ref{Hubbard_MF_H}) is determined by the following criteria.
Firstly, in the absence of impurities, homogeneous resistivity
$\rho_{0}$ should be proportional to the artificial broadening
$\eta$. Secondly, the resistivity in the case with impurities
should be proportional to the impurity
concentration. We found that a $40\times 40$ lattice is able to
achieve the above two criteria down to temperature as low as
$T_\delta=0.02$, which is roughly equal to the average energy
level spacing, and will be the system size used in the following.
Each data point is then averaged over 10 different impurity
configurations, which we found to be sufficient to ensure the
randomness of the impurity distribution. We chose
$t^{\prime}=-0.2$ and the energy unit to be $t=100$meV, which
gives temperature scale $T=0.01t\sim 10$K and the magnetic field
scale $B=0.004\sim 7$T, in the same scale as a recent study of NMR
lineshapes in the superconducting state.\cite{Harter07}

\begin{figure}[b]
\centering
\leavevmode
\begin{minipage}{.45\columnwidth}(a)
\includegraphics[width=3.5cm,height=3cm,clip=true]{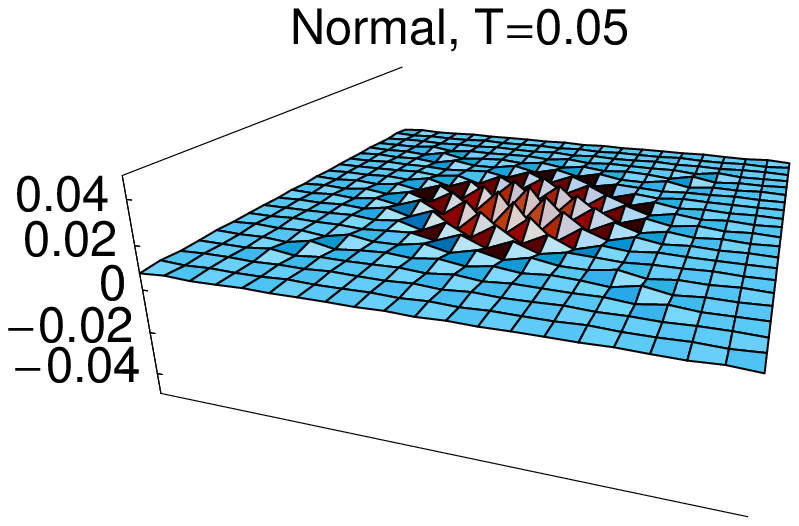}
\end{minipage}
\begin{minipage}{.45\columnwidth}(b)
\includegraphics[width=3.5cm,height=3cm,clip=true]{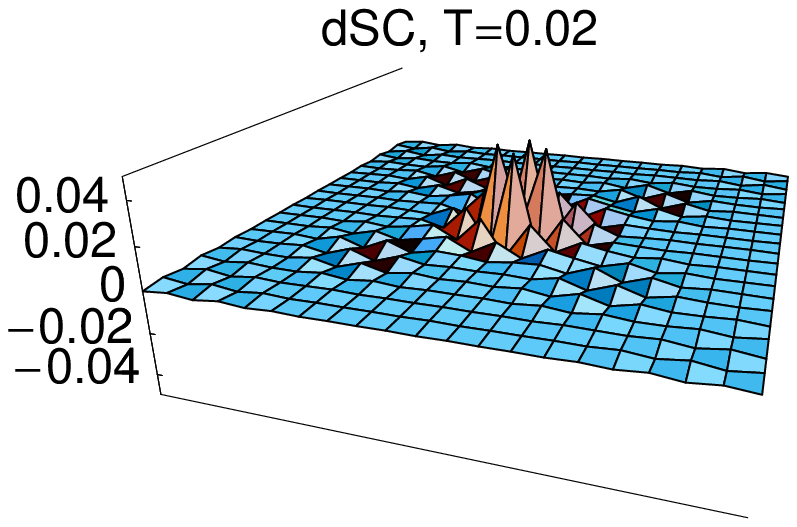}
\end{minipage}\\
\begin{minipage}{.45\columnwidth}(c)
\includegraphics[width=3.5cm,height=3.5cm,clip=true]{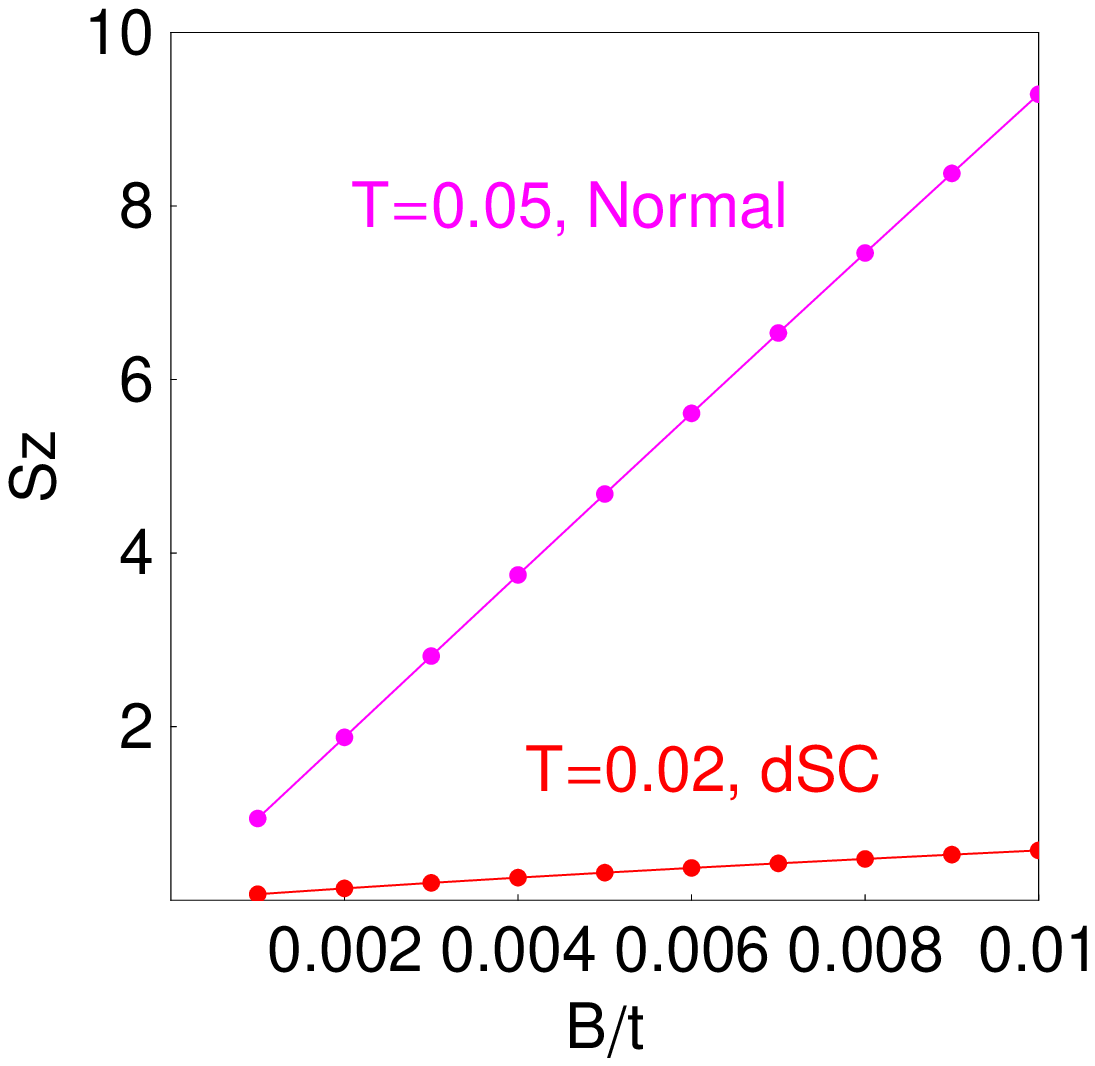}
\end{minipage}
\begin{minipage}{.45\columnwidth}(d)
\includegraphics[width=4cm,height=4cm,clip=true]{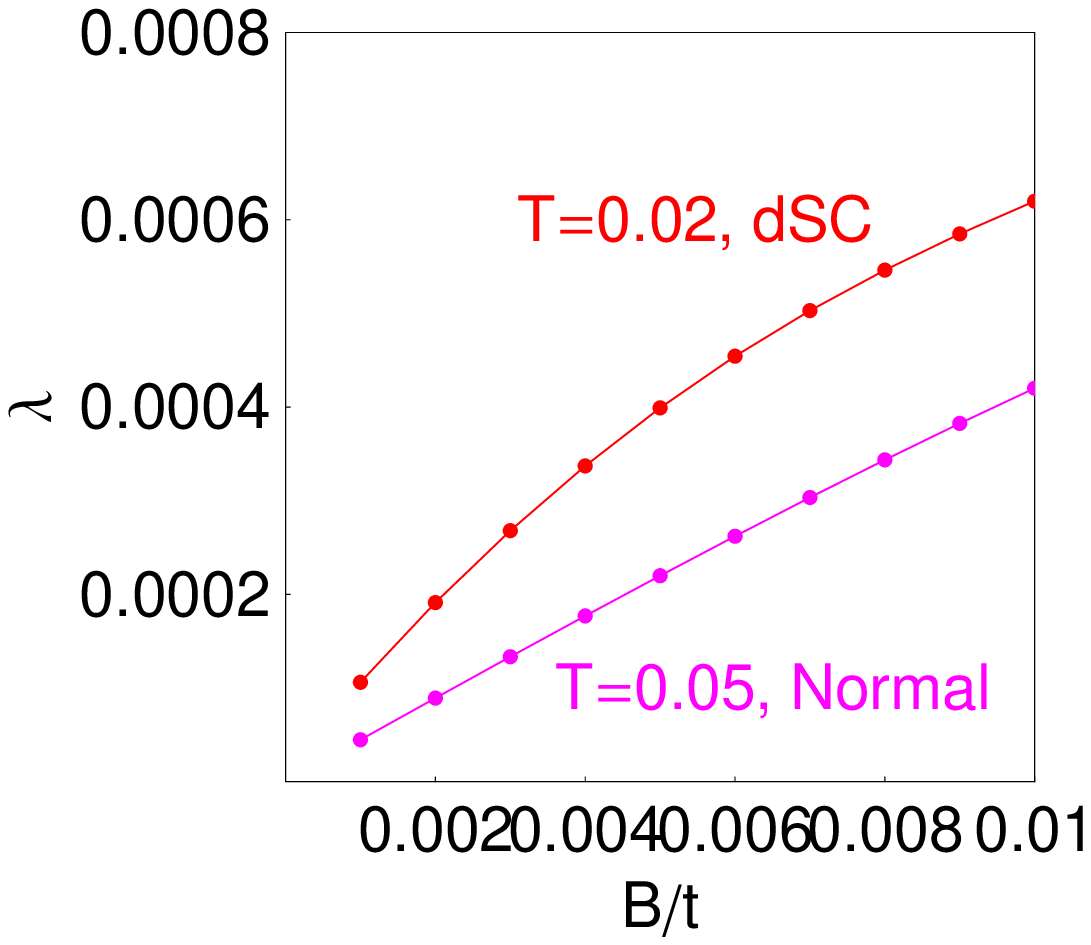}
\end{minipage}
\caption{(Color online) (a,b) Real space magnetization
pattern induced by a single nonmagnetic impurity for the normal state (a), and the dSC state (b), both at
$U=1.75$ and $B=0.01$. One sees that the dSC state has more
pronounced nearest-neighbor site  magnetization due to bound state
formation, and has smaller homogeneous magnetization due to the
opening of a gap at the Fermi surface. These effects are shown
more clearly in the total magnetization $S_{z}$ (c) and the
magnetic contrast $\lambda$ (d) versus external field. }
\label{fig:compare_SC_N_single_impurity}
\end{figure}

Before the resistivity under the influence of induced
magnetization is studied, we first compare the present study in
the normal state with the data in the $d$-wave superconducting
(dSC) state.\cite{Harter07} Such a comparison reveals the
importance of finite DOS in the normal state, as well as the bound
state formation in the dSC state. The effect of nonmagnetic
impurities in the dSC state is studied within the framework of
$d$-wave BCS theory plus magnetic correlations, equivalent to the
Hubbard model in Eq. (\ref{Hubbard_bare_H}) with additional
pairing correlations between nearest-neighbor sites $H_{pair}=\sum_{i\delta}\Delta_{\delta i}c^{\dag}_{i\uparrow}c^{\dag}_{i+\delta\downarrow}+H.c.$, where the gap is to be determined self-consistently $\Delta_{\delta i}=V\langle c_{i\uparrow}c_{i+\delta\downarrow}+c_{i+\delta\uparrow}c_{i\downarrow}\rangle/2$, with $V=1$. The real-space magnetization pattern
induced by a single nonmagnetic impurity is shown in Fig.
\ref{fig:compare_SC_N_single_impurity}(a) and Fig.
\ref{fig:compare_SC_N_single_impurity}(b), in which we found three
major differences. (1) In the presence of a magnetic field, the
normal state has homogeneous magnetization significantly larger
than that of the dSC state. This is obviously due to the opening
of the gap in the dSC state which reduces the DOS at the Fermi
level, and hence exhibits a smaller susceptibility than the normal
state. (2) The magnetization on the nearest-neighbor sites of the
impurity is drastically enhanced in the dSC state, which we found
to be consistent with the bound state formation due to the
$d$-wave symmetry.\cite{Alloul07} (3) The dSC state has a shorter
correlation length, resulting from the enhancement of
nearest-neighbor site magnetization in comparison with the
relatively smaller magnetization on the second and third nearest
sites away from the impurity. To give a quantitative description
of these features, we introduce the total magnetization $S_{z}$
and the magnetic contrast $\lambda$
\begin{eqnarray}
S_{z}=\sum_{i}m_{i}\;,\nonumber \\
\lambda=\frac{1}{N}\sum_{i}|m_{i}-m_{0}|\;,
\label{mag_order_para}
\end{eqnarray}
where $m_{i}$ is the magnetization at site $i$ and $m_{0}$ is the homogeneous magnetization in the absence of impurities but in the presence of a magnetic field. The meaning of $\lambda$ is to estimate the fluctuation of site-dependent magnetization away from its homogeneous value $m_{0}$, hence an indication of locally induced staggered moment. Since interference between impurities is always present and the local environment is different around each impurity, the deviation from $m_{0}$ of the whole system needs to be considered, and therefore we sum over $i$ for $\lambda$ in Eq. (\ref{mag_order_para}). The behavior of $S_{z}$ and $\lambda$ versus the applied field is shown in Fig. \ref{fig:compare_SC_N_single_impurity}(c) and Fig. \ref{fig:compare_SC_N_single_impurity}(d), where one sees that $S_{z}$ in the normal state is one order of magnitude larger than in the dSC state, which is attributed to the overall larger homogeneous magnetization in the normal state. However, in the $\lambda$ versus field plot, we see that after the homogeneous magnetization is subtracted, as in the definition of $\lambda$, the dSC state has a larger value due to the enhanced magnetization attributed to the bound state formation. Such a comparison indicates that DOS at the Fermi level is crucial to the formation of impurity induced moments, which in turn motivates us to propose a phenomenological model that emphasizes the effect of reducing DOS in the underdoped region, as will be discussed in Sec IV.

\section{resistivity upturns at optimal doping}

\begin{figure}[b]
\centering
\includegraphics[width=.99\columnwidth,clip=true]{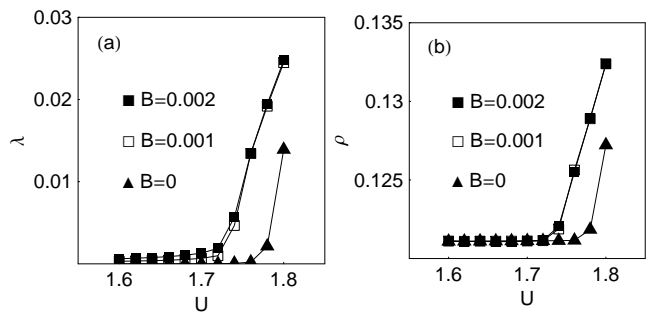}
\caption{(a) Magnetic contrast and (b) resistivity
versus $U$ at optimal doping with $T=0.03$ and 2\% impurities.}
\label{fig:optimal_scanU}
\end{figure}

Motivated by the NMR experiments,\cite{Alloul07,Ouazi04,Ouazi06,Bobroff99,Bobroff01} we study the magnetic response
in the paramagnetic region close to the magnetic phase boundary.
For convenience and direct comparison to experiments where unitary
scatterers are created by Zn substitution or irradiation defects
in YBCO, we choose $V_{imp}=100$. For a system with 2\% impurities, we
show in Fig. \ref{fig:optimal_scanU}(a) the magnetic contrast
$\lambda$ versus $U$. For the band structure used in this paper,
the critical Coulomb repulsion is found to be $U_c\sim 1.75$,
above which a spontaneous magnetization is observed for zero
field. This value is found to depend on system size and impurity
content,\cite{Harter07} but the value is roughly close to $U_c\sim
1.75$. As seen in Fig. \ref{fig:optimal_scanU}(b),the resistivity increases with $U$, and coincides with the
behavior of $\lambda$ in the region both below and above its
critical value. This  positive correlation between $\lambda$ and
$\rho$ serves as the first evidence that we can attribute the
increase of resistivity to the extra scattering induced by the
magnetic moments. In the following discussion we choose $U=1.74$
such that it is close to but slightly below the critical $U_c$,
and the system exhibits paramagnetic response to an external
field.

We note that in the region where the system cross the
magnetic phase boundary, for instance at large $U$ or low
temperatures, numerics found that there are several stable states
with comparable energies competing with each other. Taking
different initial conditions or a different route for the
convergence can result in a different apparent ground state
configuration; for instance, we found a charge density wave (CDW)
ground state with periodicity $(\pi,\pi)$ that can exist in large
U and zero field, consistent with the spin or charge modulated
state found in other studies with a sufficiently large Coulomb
repulsion.\cite{Andersen07,Robertson06,delMaestro06,Atkinson05,Vojta06}
However, considering the strong experimental evidence of magnetic
ordering, as well as the resulting resistivity in comparison with
the transport measurement, only the paramagnetic induced moment
state can give a proper description of both induced magnetization
and transport anomalies in the optimal to lightly underdoped
systems, and hence will be the stable configuration focused on in
this report.

\begin{figure}[t]
\centering
\includegraphics[width=0.99\columnwidth,clip=true]{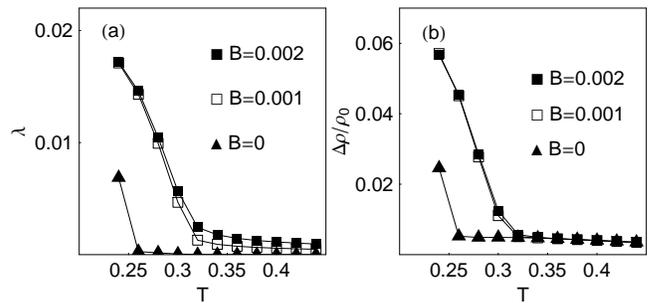}
\caption{(a) Magnetic contrast $\lambda$ and (b) change of resistivity $\Delta\rho/\rho_{0}$ versus $T$ at optimal doping with $U=1.74$ and 2\% impurities. }
\label{fig:optimal_scanT}
\end{figure}


Due to the limited system size, we are unable to explore the
extremely low $T$ regime, which prevents us from comparing the
present theory with the experimentally observed Log-$T$ divergence.
However, numerics down to as low as $T=0.026\sim 26K$ shows
significant resistivity upturns in comparison with the zero field
case. Figure \ref{fig:optimal_scanT} shows both magnetic contrast
and change of resistivity $\Delta\rho/\rho_{0}$ versus temperature
$T$, where $\rho_{0}$ is the resistivity at the uncorrelated zero
field case($U=0$,$B=0$), and one sees again the positive
correlation between these two quantities. The lowest temperature
explored is slightly lower than the critical temperature
$T_{spon}\sim 0.025$ below which a spontaneous magnetization is
observed in the zero field. The magnitude of the upturn at
$T=0.026$ in comparison with high temperature resistivity is  of
the order of 5\%, roughly consistent with the value obtained in
slightly underdoped YBCO after the linear-$T$
contribution has been subtracted.\cite{Segawa99}

\begin{figure}[b]
\centering
\includegraphics[width=.99\columnwidth,clip=true]{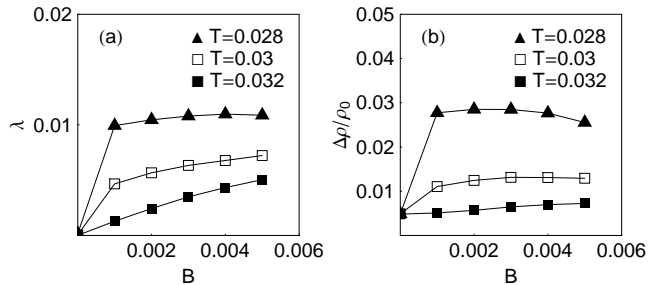}
\caption{(a) Magnetic contrast and (b) change of
resistivity versus $B$ at  optimal doping with $U=1.74$ and 2\%
impurities. } \label{fig:optimal_scanB}
\end{figure}

The magnetoresistance in the presence of induced magnetization is
shown in Fig. \ref{fig:optimal_scanB}, where we again see a
positive correlation between  $\lambda$ and $\Delta\rho/\rho_{0}$
with increasing magnetic field $B$. At the temperatures where the
resistivity upturns set in, we found that both $\lambda$ and
$\Delta\rho/\rho_{0}$ first increase with the field, and
eventually saturate and slightly decrease in the high field
region. One can unambiguously define a field scale $B_{sat}$ above
which $\lambda$ and $\Delta\rho/\rho_{0}$ saturate, and we found
that $B_{sat}$ decreases as temperature is lowered. Such a
increase-saturation behavior is consistent with the
magnetoresistance observed in YBCO,\cite{Rullier-Albenque07a,Rullier-Albenque07b}
although $B_{sat}$ observed therein is slightly higher, possibly
due to the higher field required to eliminate the
superconductivity before normal state properties can be observed.
Since $B_{sat}$ decreases as lowering temperatures, the region
where the magnetic contrast $\lambda$ is linear with respect to
the external field also decreases accordingly, which indicates
that as the magnetization starts to grow at low temperatures, the
interference between the magnetic islands induced around each
impurity is also enhanced, causing $\lambda$ to deviate from a
linear response.

\begin{figure}[b]
\centering
\includegraphics[width=0.99\columnwidth,clip=true]{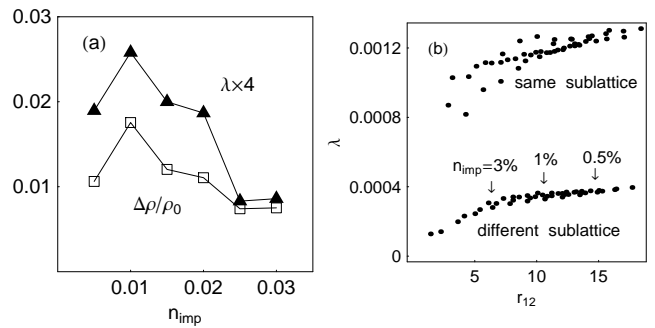}
\caption{(a) Magnetic contrast $\lambda$ and change
of resistivity $\Delta\rho/\rho_{0}$ versus $n_{imp}$ at optimal
doping with $U=1.74$, $B=0.001$, $T=0.03$, and (b)$\lambda$
induced by the 2-impurity model plotted against the seperation
between the two impurities $r_{12}$, collecting all relative
positions up to thirdteenth shell. Values of $r_{12}$ that correspond to average distance of impurities at $n_{imp}=3\%$, $1\%$, and $0.5\%$ are indicated.  } \label{fig:optimal_scannimp}
\end{figure}

The last issue we need to address is the behavior of $\lambda$ and
$\Delta\rho/\rho_{0}$ as changing impurity concentration
$n_{imp}$, in comparison with the available experimental data
which shows that the resistivity upturns monotonically increase
with $n_{imp}$ up to $n_{imp}\sim 3\%$. Fig.
\ref{fig:optimal_scannimp} shows the numerical result under the
influence of changing $n_{imp}$, where one again sees the
consistency between the behavior of $\lambda$ and
$\Delta\rho/\rho_{0}$. However, instead of increasing
monotonically with increasing $n_{imp}$, we found that both
$\lambda$ and $\Delta\rho/\rho_{0}$ increase up to a critical
concentration $n^{c}_{imp}\sim 1\%$, and then decrease as more
impurities are introduced on the plane. Such a result indicates
that the impurity induced magnetization is proportional to
$n_{imp}$ only up to a certain extend, beyond which the
interference takes place and eventually destroys the magnetization
and the associated magnetic scattering. To further demonstrate
that the interference effect is more destructive than constructive
to the induced magnetization, we study the 2-impurity case in the
present model, and plot $\lambda$ against the separation between
the two impurities, as shown in Fig.
\ref{fig:optimal_scannimp}(b). We first found that there exists a
strong enhancement of magnetization if both impurities are on the
same sublattice, consistent with previous studies in the dSC
state.\cite{Chen04,Andersen07} Secondly, $\lambda$ indeed
decreases as the two impurities get closer, which is the case when
$n_{imp}$ is increased, indicating the destructive nature of the
interference effect, and hence the decreasing of magnetization at
sufficiently large impurity content. Our result therefore predicts
that if the extremely disordered samples ($n_{imp}>3\%$) can be
studied experimentally, a critical concentration can occur beyond
which the resistivity upturn drops as increasing impurity content,
assuming that weak localization has not yet taken place. The
critical concentration $n_{imp}\sim 1\%$ shown in the present
study is apparently smaller than the experimental value, which may
be due to a smaller linear response region in the present model in comparison with the real cuprates,
presumably an artifact of such a weak coupling mean field approach. In addition, the critical disorder concentration $n_{imp}$ will depend on the details of the disorder modeling, for instance the nature of the disorder, or the extent of the impurity potential, which is outside of the scope of our study.

\section{Effect of pseudogap in DOS on Resistivity }

From a weak coupling perspective, we expect enhancement of
resistivity upturns as the system  is underdoped, based on the
following two features: Firstly, correlations are more prominent
as one goes toward half-filling, resulting in an increase of the
effective $U$ entering our model. Although the
Hartree-Fock type mean field theory can not capture the Mott
transition induced by correlations nor the pseudogap phenomenon,
the drastic increase of resistivity near the critical value of $U$
suggests that correlations indeed affect resistivity as one
approaches the strong coupling region. The large $U$ region in
Fig. \ref{fig:optimal_scanU} demonstrates that correlation
strength $U$, as well as the induced magnetic moment, are indeed
essential ingredients to determine the magnitude of the upturn.

\begin{figure}[ht!]
\centering \leavevmode
\begin{minipage}{1.0\columnwidth}
\includegraphics[width=0.8\columnwidth,clip=true]{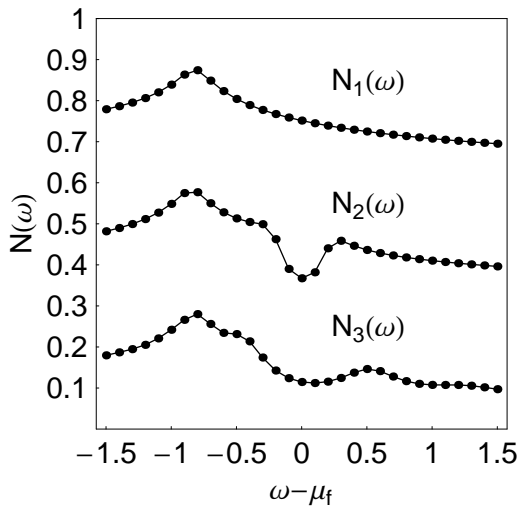}
\end{minipage}
\caption{Comparison of DOS: the normal state dispersion $\xi_{k}$
gives $N_{1}(\omega)$ (shifted), the proposed phenomenological
model for the pseudogap state $E_{k}$ gives
$N_{2}(\omega)$ (shifted), and the actual $N_{3}(\omega)$(original
scale) given by the effective hopping model after Fourier
transform of $E_{k}$ in a $40\times 40$ system, with
$\mu_{f}=0.02$ and $\Delta=0.2$.} \label{fig:PG_DOS}
\end{figure}

Secondly, the opening of the pseudogap in the quasiparticle
spectrum is known to favor bound state formation, which  in turn
promotes the impurity induced magnetic moment.\cite{Alloul07} This
is similar to the dSC state where the pole of
impurity T-matrix falls within the gap, producing a bound state
localized around the impurity. We expect that the reduction of the
DOS in the pseudogap state also produces poles of T-matrix near
Fermi energy, although the exact form of Green's function and
Dyson's equation remains unknown. Resistivity upturns are then
affected by the pseudogap formation, based on the naive argument
that impurity induced moments result in the upturn. To get a crude
idea of the effect of reducing the  DOS, we introduce a pseudogap
in an {\em ad hoc} way without going through the T-matrix
formalism, since no microscopic model of the pseudogap state is
generally agreed upon at present. The following form of dispersion
and DOS is proposed for the homogeneous pseudogap state
\begin{eqnarray}
E_{k}&=&\mbox{sign}(\xi_{k})\sqrt{\xi^{2}_{k}+\Delta^{2}_{k}}\;, \nonumber\\
N(\omega)&=&\int \frac{dk^{2}}{4\pi^{2}} \frac{\eta/\pi}{(\omega-E_{k})^{2}-\eta^{2}}\;,
\label{PG_DOS_model_1}
\end{eqnarray}
where
$\xi_{k}=-2t(\cos(k_{x})+\cos(k_{y}))-4t^{\prime}\cos(k_{x})\cos(k_{y})-\mu_{f}$
is the  normal metallic dispersion, with a constant ``pseudogap"
$\Delta_{k}=0.2$. We then Fourier transform $E_{k}$ back to real
space and find an effective long range hopping model that gives
the energies $E_{k}$. The hopping amplitude $t_{ij}$ of this
extended hopping model is therefore
\begin{equation}
t_{ij}=\int \frac{dk^{2}}{4\pi^{2}} E_{k}\{\cos[k_{x}\cdot(x_{i}-x_{j})]+\cos[k_{y}\cdot(y_{i}-y_{j})]\}\;.
\label{PG_DOS_model_2}
\end{equation}

We calculate the hopping range up to $|x_{i}-x_{j}|=|y_{i}-y_{j}|=20$ on a $40\times 40$ lattice. Numerics show a roughly $40\%$ reduction of DOS at the chemical potential, as shown in Fig. \ref{fig:PG_DOS}. The calculation of resistivity then follows Eq. (\ref{tight_binding_current}) and (\ref{jj_cor_fn}), while the contribution from all hopping terms $t_{\delta}=t_{ij}$ and their corresponding distance $\vec{\delta}=\vec{r}_{i}-\vec{r}_{j}$ all need to be considered.

\begin{figure}[ht!]
\centering
\includegraphics[width=.99\columnwidth,clip=true]{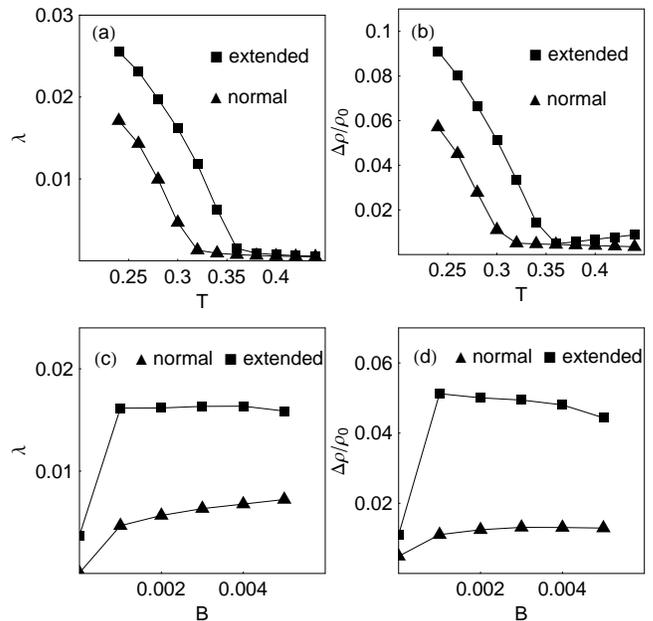}
\caption{Comparison of (a) $\lambda$ and
(b) $\Delta\rho/\rho_{0}$ for models with (Extended) and
without (Normal) reduction of DOS by applying extended hopping Eq.
(\ref{PG_DOS_model_2}), both at optimal doping, with $U=1.75$,
$B=0.001$ and 2\% impurities. (c) and (d): same quantities vs. $B$
for Normal and Extended hopping models.} \label{fig:PG_scanTB}
\end{figure}

Figure \ref{fig:PG_scanTB} (a,b) shows the magnetization and
resistivity comparing extended hopping model with the normal state
Hubbard model Eq. (\ref{Hubbard_bare_H}) which contains only
nearest and next-nearest neighbor hopping. We fix both models at
optimal doping $\delta=0.15$ and examine solely the effect of
reducing DOS. Among the magnetic field region explored
$0<g\mu_{B}B/2<0.01$, the magnetic contrast $\lambda$ is found to
be enhanced in the extended hopping model, confirming our hypothesis
of reducing DOS promotes bound state formation, which also gives
slightly larger resistance between temperature range
$0.02<T<0.045$. The magnetization and resistivity versus field is
shown in Fig. \ref{fig:PG_scanTB}(c,d), where one sees larger
magnetization comparing to the normal state model, with a smaller
linear response region and the saturation at high field is again
revealed. Resistivity upturns are enhanced overall in both low and
high field region, and is consistent with the behavior of
$\lambda$. The hypothesis of reducing DOS promotes induced
moments, and in turn enhances the resistivity upturns, is then
well proved.


\section{Conclusions}

In summary, we employed a Hartree-Fock decomposition of the
Hubbard model to study transport properties under the influence of
disorder induced magnetization, which is a consequence of the
interplay between strong correlations and inhomogeneity. The
numerical results suggest that, at low enough temperatures and
strong enough correlations,  impurity induced magnetization is
drastically enhanced. Within this regime, both induced
magnetization and resistivity are increased as (1) the temperature
is lowered, (2) the magnetic correlations are enhanced, (3)  the
magnetic field is increased, and (4) more impurities are
introduced, consistent with the conditions in which the
enhancement of resistivity is observed experimentally. We predict,
in addition, that the addition of further disorder can sometimes
lead to a nonmonotonic field dependence as the magnetic potential
landscape becomes smooth; this property has not yet been observed
to our knowledge. Extremely heavily disordered or strongly
correlated samples will lie in a different regime, which we have
not yet treated, where disorder will create a spontaneous,
short-range ordered magnetic state even in zero
field\cite{Andersen07}; in this case we anticipate that the
magnetoresistance will quite small.

The positive correlation between induced magnetization and
resistivity confirms our hypothesis that the enlarged cross
section due to these local magnetic moments gives extra scattering
and hence the resistivity upturns, and indicates that the
hole-doped cuprates lie within this regime over a wide range of
(under) doping, in which strong correlations can cause anomalies
in the thermodynamic observables. A phenomenological model that
produces reduction of DOS near the Fermi level in an {\it ad hoc}
way further suggests that, as the system is underdoped, besides
the enhancement of correlations that can increase the resistivity,
the anomalous energy spectrum in the underdoped region can promote
the impurity bound state and hence the magnetization, which in
turn boosts the magnetic scattering and the resistivity upturns.
The proposed mean field theory plus real space diagonalization
scheme is therefore a powerful tool to capture the complex effect
on the transport properties due to strong correlations,
inhomogeneity, and the spectral anomalies in the low temperature
region where the transport is dominated by disorder.   Further
applications of the present theory, as well as the influence of
impurity induced magnetization on other thermodynamic observables
in the metallic cuprates, will be addressed in a future study.

\section{Acknowledgments}

We appreciate useful discussions with H. Alloul, P. Fournier, L.
Taillefer, and F. Rullier-Albenque regarding the interpretation of
experiments, and O. P. Sushkov, H. Kontani, D. Maslov, J. Harter,
A. T. Dorsey, K. Ingersent, and M. Gabay for various theoretical
aspects. Partial funding for this research was provided by DOE-BES
DE-FG02-05ER46236. B. M. A. acknowledges support from the Villum
Kann Rasmussen foundation.


\begin{thebibliography}{999}

\bibitem{Chien91}
T. R. Chien, Z. Z. Wang, and N. P. Ong, Phys. Rev. Lett. {\bf 67}, 2088 (1991).
%
\bibitem{Ito93}
T. Ito, K. Takenaka, and S. Uchida, Phys. Rev. Lett. {\bf 70}, 3995 (1993).
%
\bibitem{Ando95}
Y. Ando, G. S. Boebinger, A. Passner, T. Kimura, and K. Kishio, Phys. Rev. Lett. {\bf 75}, 4662 (1995).
%
\bibitem{Boebinger96}
G. S. Boebinger, Y. Ando, A. Passner, T. Kimura, M. Okuya, J. Shimoyama, K. Kishio, K. Tamasaku, N. Ichikawa, and S. Uchida, Phys. Rev. Lett. {\bf 77}, 5417 (1996).

%
\bibitem{Hanaki01}
Y. Hanaki, Y. Ando, S. Ono, and J. Takeya, Phys. Rev. B {\bf 64}, 172514 (2001).
%
\bibitem{Ono00}
S. Ono, Y. Ando, T. Murayama, F. F. Balakirev, J. B. Betts, and G. S. Boebinger, Phys. Rev. Lett. {\bf 85}, 638 (2000).
%
\bibitem{Fukuzumi96}
Y. Fukuzumi, K. Mizuhashi, K. Takenaka, and S. Uchida, Phys. Rev. Lett. {\bf 76}, 684 (1996).
%
\bibitem{Segawa99}
K. Segawa and Y. Ando, Phys. Rev. B {\bf 59}, R3948 (1999).
%
\bibitem{Segawa01}
K. Segawa and Y. Ando, Phys. Rev. Lett. {\bf 86}, 4907 (2001).
%
\bibitem{Walker95}
D. J. C. Walker, A. P. Mackenzie, and J. R. Cooper, Phys. Rev. B {\bf 51}, 15653 (1995).
%
\bibitem{Rullier-Albenque00}
F. Rullier-Albenque, P. A. Vieillefond, H. Alloul, A. W. Tyler, P. Lejay, and J. F. Marucco, Eur. Phys. Lett. {\bf 50}, 81 (2000).
%
\bibitem{Rullier-Albenque01}
F. Rullier-Albenque, H. Alloul, and R. Tourbot, Phys. Rev. Lett. {\bf 87}, 157001 (2001).
%
\bibitem{Rullier-Albenque03}
F. Rullier-Albenque, H. Alloul, and R. Tourbot, Phys. Rev. Lett. {\bf 91}, 047001 (2003).
%
\bibitem{Rullier-Albenque06}
F. Rullier-Albenque, R. Tourbot, H. Alloul, P. Lejay, D. Colson, and A. Forget, Phys. Rev. Lett. {\bf 96}, 067002 (2006).
%
\bibitem{Rullier-Albenque07a}
F. Rullier-Albenque, H. Alloul, F. Balakirev, and C. Proust, Eur. Phys. Lett. {\bf 81}, 37008 (2008).
%
\bibitem{Rullier-Albenque07b}
F. Rullier-Albenque, H. Alloul, C. Proust, P. Lejay, A. Forget, and D. Colson, Phys. Rev. Lett. {\bf 99}, 027003 (2007).
%
\bibitem{Alloul07}
H. Alloul, J. Bobroff, M. Gabay, and P. J. Hirschfeld, Rev. Mod.
Phys. {\bf 81}, 45 (2009).
%
\bibitem{Lake01}
B. Lake, G. Aeppli, K. N. Clausen, D. F. McMorrow, K. Lefmann, N. E. Hussey, N. Mangkorntong, M. Nohara, H. Takagi, T. E. Mason, and A. Schroder, Science {\bf 291}, 1759 (2001).
%
\bibitem{Lake02}
B. Lake, H. M. R\o nnow, N. B. Christensen, G. Aeppli, K. Lefmann, D. F. McMorrow, P. Vorderwisch, P. Smeibidl, N. Mangkorntong, T. Sasagawa, M. Nohara, H. Takagi, and T. E. Mason, Nature {\bf 415}, 299 (2002).
%
\bibitem{Katano00}
S. Katano, M. Sato, K. Yamada, T. Suzuki, and T. Fukase, Phys. Rev. B {\bf 62}, R14677 (2000).
%
\bibitem{Khaykovich02}
B. Khaykovich, Y. S. Lee, R. W. Erwin, S.-H. Lee, S. Wakimoto, K. J. Thomas, M. A. Kastner, and R. J. Birgeneau, Phys. Rev. B {\bf 66}, 014528 (2002).
%
\bibitem{HKimura:2003} H. Kimura, M. Kofu, Y. Matsumoto, and K. Hirota, Phys. Rev. Lett. {\bf 91}, 067002 (2003).
%
\bibitem{Suzuki98}
T. Suzuki, T. Goto, K. Chiba, T. Shinoda, T. Fukase, H. Kimura, K. Yamada, M. Ohashi, and Y. Yamaguchi, Phys. Rev. B {\bf 57}, R3229 (1998).
%
\bibitem{Wakimoto01}
S. Wakimoto, R. J. Birgeneau, Y. S. Lee, and G. Shirane, Phys. Rev. B {\bf 63}, 172501 (2001).
%
\bibitem{Haug09}
D. Haug, V. Hinkov, A. Suchaneck, D. S. Inosov, N. B. Christensen, Ch. Niedermayer, P. Bourges, Y. Sidis, J. T. Park, A. Ivanov, C. T. Lin, J. Mesot, and B. Keimer, arXiv:0902.3335v1.
%

%
\bibitem{Ouazi04}
S. Ouazi, J. Bobroff, H. Alloul, and W. A. MacFarlane, Phys. Rev. B {\bf 70}, 104515 (2004).
%
\bibitem{Ouazi06}
S. Ouazi, J. Bobroff, H. Alloul, M. Le Tacon, N. Blanchard, G. Collin, M. H. Julien, M. Horvatic, and C. Berthier, Phys. Rev. Lett. {\bf 96}, 127005 (2006).
%
\bibitem{Bobroff99}
J. Bobroff, W. A. MacFarlane, H. Alloul, P. Mendels, N. Blanchard, G. Collin, and J.-F. Marucco, Phys. Rev. Lett. {\bf 83}, 4381 (1999).
%
\bibitem{Bobroff01}
J. Bobroff, H. Alloul, W. A. MacFarlane, P. Mendels, N. Blanchard, G. Collin, and J.-F. Marucco, Phys. Rev. Lett. {\bf 86}, 4116 (2001).
%
\bibitem{ChNiedermayer:1998} Ch. Niedermayer, C. Bernhard, T. Blasius, A. Golnik, A. Moodenbaugh, and J. I. Budnick, Phys. Rev. Lett. {\bf 80}, 3843 (1998).
%
\bibitem{CPanagopoulos:2002} C. Panagopoulos, J. L. Tallon, B. D. Rainford, T. Xiang, J. R. Cooper, and C. A. Scott, Phys. Rev. B {\bf 66},
064501 (2002).
%
\bibitem{TAdachi:2004}T. Adachi, S. Yairi, K. Takahashi,  Y.
Koike, I. Watanabe, and  K. Nagamine, Phys. Rev. B {\bf 69}, 184507 (2004).
%
\bibitem{CPanagopoulos:2005} C. Panagopoulos and V. Dobrosavljevi\'{c}, Phys. Rev. B {\bf 72},
014536 (2005).
%
\bibitem{SSanna:2004} S. Sanna, G. Allodi, G. Concas, A. D. Hillier, and R. De Renzi, Phys. Rev. Lett. {\bf 93}, 207001 (2004).
%
\bibitem{RIMiller:2006} R. I. Miller, R. F. Kiefl, J. H. Brewer, Z. Salman, J. E. Sonier, F. Callaghan, D. A. Bonn, W. N. Hardy, and R. Liang, Physica B {\bf 374-375}, 215 (2006).
%
\bibitem{Altshuler79}
B. L. Altshuler and A. G. Aronov, Solid State Comm. {\bf 36}, 115 (1979).
%
\bibitem{Altshuler80}
B. L. Altshuler, A. G. Aronov, and P. A. Lee, Phys. Rev. Lett. {\bf 44}, 1288 (1980).
%
\bibitem{PALee85}
P. A. Lee and T. V. Ramakrishnan, Rev. Mod. Phys. {\bf 57}, 287 (1985).

\bibitem{Fournier00} P. Fournier, J. Higgins, H. Balci, E. Maiser,
C.J. Lobb, and R.L. Greene, Phys. Rev. B 62, R11993 (2000).
%
\bibitem{Beloborodov07}
I. S. Beloborodov, A. V. Lopatin, V. M. Vinokur, and K. B. Efetov, Rev. Mod. Phys. {\bf 79}, 469 (2007).
%
\bibitem{Kontani99}
H. Kontani, K. Kanki, and K. Ueda, Phys. Rev. B {\bf 59}, 14723 (1999).
%
\bibitem{Kontani06}
H. Kontani and M. Ohno, Phys. Rev. B {\bf 74}, 014406 (2006).
%
\bibitem{Kontani08}
H. Kontani, Rep. Prog. Phys. {\bf 71}, 026501 (2008).
%
\bibitem{Alvarez05}
G. Alvarez,  M. Mayr, A. Moreo, and E. Dagotto, Phys. Rev. B {\bf 71}, 014514 (2005); M. Mayr,
G. Alvarez, A. Moreo, and E. Dagotto, Phys. Rev. B {\bf 73}, 014509 (2006).
%
\bibitem{Andersen07}
B. M. Andersen, P. J. Hirschfeld, A. P. Kampf, and M. Schmid, Phys. Rev. Lett. {\bf 99}, 147002 (2007).
%
\bibitem{Atkinson07}
W. A. Atkinson, Phys. Rev. B {\bf 75}, 024510 (2007).
%
\bibitem{Andersen06}
B. M. Andersen and P. J. Hirschfeld, Physica C (Amsterdam) {\bf 460-462}, 744 (2007).
%
\bibitem{Andersen08}
B. M. Andersen and P. J. Hirschfeld, Phys. Rev. Lett. {\bf 100}, 257003 (2008).
%
\bibitem{Alvarez08}
G. Alvarez and E. Dagotto, Phys. Rev. Lett. {\bf 101}, 177001 (2008).
%
\bibitem{Takigawa02}
M. Takigawa, M. Ichioka, and K. Machida, Eur. Phys. J. B {\bf 27}, 303 (2002).
%
\bibitem{Harter07}
J. W. Harter, B. M. Andersen, J. Bobroff, M. Gabay, and P. J. Hirschfeld, Phys. Rev. B {\bf 75}, 054520 (2007).
%
\bibitem{Robertson06}
J. A. Robertson, S. A. Kivelson, E. Fradkin, A. C. Fang, and A. Kapitulnik, Phys. Rev. B {\bf 74}, 134507 (2006).
%
\bibitem{delMaestro06}
A. Del Maestro, B. Rosenow, and S. Sachdev, Phys. Rev. B {\bf 74}, 024520 (2006).
%
\bibitem{Atkinson05} W. A. Atkinson, Phys. Rev. B {\bf 71}, 024516 (2005).
%
\bibitem{Vojta06}
M. Vojta, T. Vojta, and R. K. Kaul, Phys. Rev. Lett. {\bf 97}, 097001 (2006).
%
\bibitem{Chen04}
Y. Chen and C. S. Ting, Phys. Rev. Lett. {\bf 92}, 077203 (2004).
%


\end{thebibliography}
\end{document}